\documentclass[article]{aa}

\usepackage{longtable}
\usepackage{graphicx,caption}
\usepackage[normalem]{ulem}

\usepackage{color}

\usepackage{babel}
\usepackage{enumitem}
\usepackage[varg]{txfonts}

\usepackage{natbib}
\usepackage{ulem}

\usepackage{caption}

\definecolor{pad}{RGB}{242,12,242}

\def\degr{\hbox{$^\circ$}}

\makeatother

\begin{document}

\title{Note on the dynamical evolution of C/2017 K2 PANSTARRS}

\author{Małgorzata Królikowska\inst{\ref{inst1}} and Piotr A. Dybczyński\inst{\ref{inst2}} }

\institute{
Space Research Centre of Polish Academy of Sciences, Bartycka 18A, Warszawa, Poland, \email{mkr@cbk.waw.pl} \label{inst1}
\and Astronomical Observatory Institute, Faculty of Physics, A.Mickiewicz University, Słoneczna 36, Poznań, Poland, 
\email{dybol@amu.edu.pl}\label{inst2}
}

\date{}

\abstract{Comet C/2017 K2 PANSTARRS drew attention to its activity already at a time of its  discovery in May 2017 when it was about 16\,au from the Sun. This Oort spike comet will approach its perihelion in December 2022, and the question about its dynamical past is one of the important issues to explore.}{In order to answer the question whether  C/2017 K2 is a dynamically old or new comet it is necessary to obtain its precise osculating orbit, its original orbit, and propagate its motion backwards in time to the previous perihelion. The knowledge of the previous perihelion distance is necessary to distinguish between these two groups of the Oort spike comets.}{We study a dynamical evolution of C/2017 K2 to the previous perihelion (backward calculations for about 3-4\,Myr) as well as to the future (forward calculations for about 0.033\,Myr) using the swarm of virtual comets (VCs) constructed from a nominal osculating orbit of this comet which we determined here using all positional measurements available at the moment. Outside the planetary system both Galactic and stellar perturbations were taken into account.}{We derived that C/2017~K2 is a dynamically old Oort spike comet ($1/a_{\rm prev}=(48.7\pm 7.9)\times 10^{-6}$\,au$^{-1}$) with the previous perihelion distance below 10\,au for 97 per cent of VCs (nominal $q_{\rm prev}=3.77$\,au).  According to the present data this comet will be perturbed into a more tightly bound orbit after passing the planetary zone ($1/a_{\rm next}=(1140.2\pm 8.0)\times 10^{-6}$\,au$^{-1}$, $q_{\rm next}=1.79334\pm 0.00006$\,au) provided that non-gravitational effects will not change the orbit significantly.}{C/2017 K2 has already visited our planetary zone during its previous perihelion passage. Thus, it is almost certainly a dynamically old Oort spike comet. The future orbital solution of this comet is formally very precise, however, it is much less definitive since the presented analysis is based on pre-perihelion data taken at very large heliocentric distances (23.7--14.6 au from the Sun), and this comet can experience a significant non-gravitational perturbation during the upcoming perihelion passage in 2022.}


\titlerunning{Dynamical evolution of C/2017 K2 PANSTARRS}

\authorrunning{M.Królikowska \& P.A.Dybczyński}

\maketitle

\section{Introduction}

\begin{figure}
	\includegraphics[width=8.8cm]{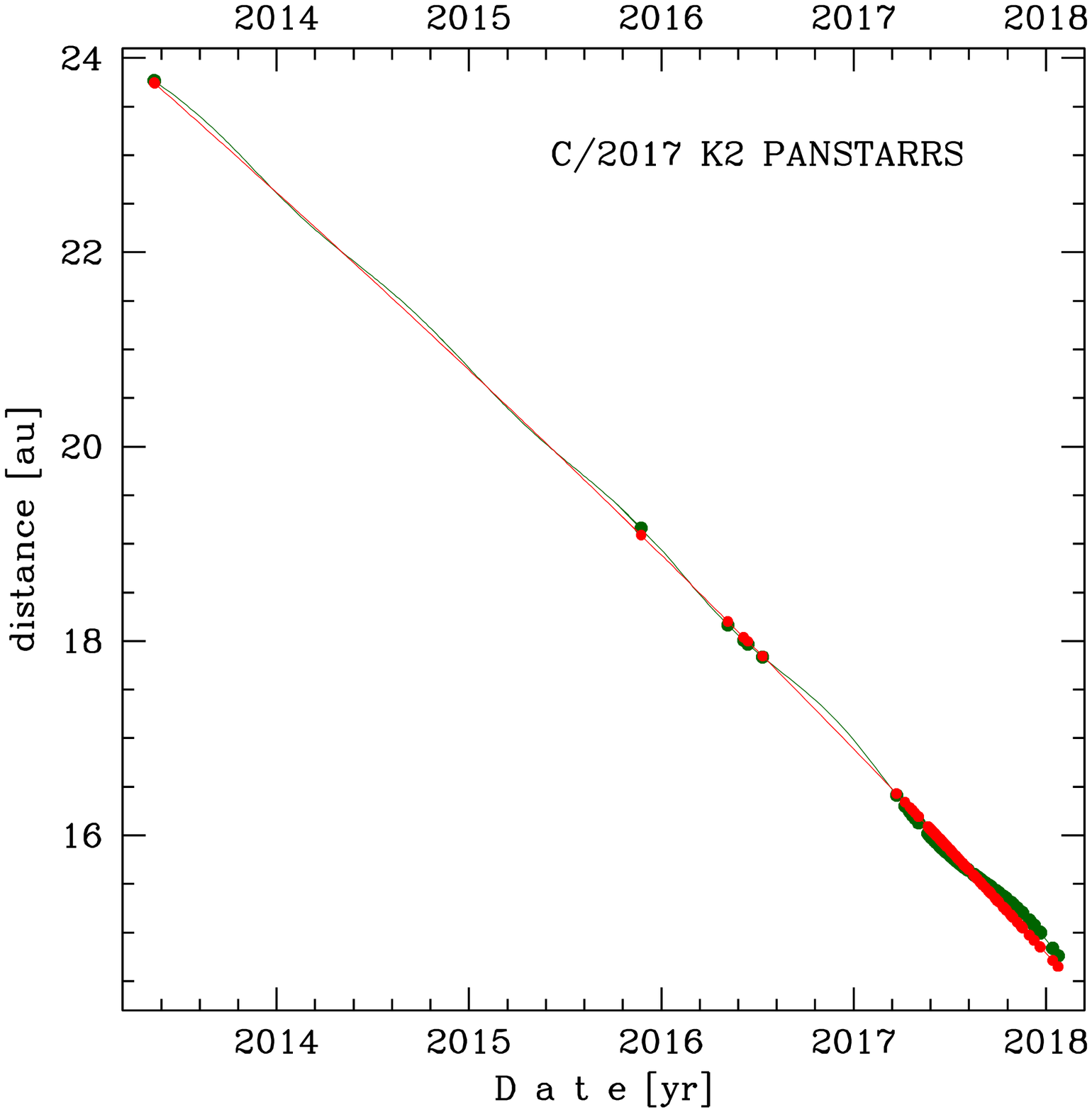} 
\protect\caption{\label{fig:data_dist} Heliocentric (thin red curve) and geocentric (green curve) distance of K2 during the time interval given in Table~\ref{tab:orbit_observed}. Time distribution of positional observations of K2 with corresponding heliocentric distance at which they were taken is shown  as red dots. The geocentric distances of measurements are given as green dots but they are almost covered by red dots due to similarly large distances from the Sun and the Earth.}

\end{figure}

This short analysis was  inspired by a very interesting study of the activity and past dynamical motion of comet  C/2017~K2 PANSTARRS recently published by \citet{hui:2017}. Here, we focus on dynamical aspects of their paper. The authors examined a history of K2\footnote{We also use  'K2' in short for C/2017~K2, as \citet{hui:2017} did} for  1\,Myr taking into account planetary perturbations, relativistic corrections and perturbations  from a Galactic tide. At the epoch of 1\,Myr in the past they obtained $1/a\simeq{(36.1\pm 17.1)} \,\times 10^{-6}$\,au$^{-1}$ and using this value alone they tried  to deduce whether a comet is dynamically new or old. This value of $1/a$ places K2 in an area of the  Oort spike region where dynamically new and dynamically old comets coexist \citep{kroli_dyb:2017}. Therefore, \citet{hui:2017} concluded that the dynamical status of K2 'must be regarded as unsettled' since they stopped the integration more or less in the middle of the time interval to the previous perihelion.

This conclusion drew our attention to K2. In the following, we discuss its dynamical evolution  backwards to the previous perihelion and forwards to the next one.

\begin{table*}
\caption{Osculating heliocentric orbit of K2 based on 450 positional measurements spanning the time interval from 2013 May  12 to 2018 January 23 available at MPC on 21 February 2017. Equator and ecliptic of J2000 is used. Solution A1 is based on weighted measurements whereas A0 is based on unweighted data. For comparison, JPL-solution is presented in the last column; it is based on 446 observations and the same interval of data (retrieved from JPL Small-Body Database Browser on 21 February 2018). }
\label{tab:orbit_observed}
{\small{
\setlength{\tabcolsep}{6.5pt} 
\begin{tabular}{lccc}
\hline 
\hline
 &  &  & \\
                                           &  Solution  A1                 & Solution A0                  & JPL solution \\
 &  &  & \\
\hline 
 &  &  & \\
perihelion distance [au]                   & 1.81086204 $\pm$ 0.00005500        & 1.81081861 $\pm$ 0.00006434        & 1.81072035 $\pm$ 0.00013422 \\
eccentricity                               & 1.00032653 $\pm$ 0.00001443        & 1.00031764 $\pm$ 0.00001823        & 1.00029212 $\pm$ 0.00003797 \\
time of perihelion passage [TT]            & 2022\,12\,21.412478 $\pm$ 0.030195 & 2022\,12\,21.420049 $\pm$ 0.041069 & 2022\,12\,21.465966 $\pm$ 0.085538 \\
inclination [deg]                          &  87.553729 $\pm$  0.000023         &  87.553662 $\pm$  0.000045         &  87.553645  $\pm$ 0.000095 \\
argument of perihelion [deg]               & 236.021084 $\pm$  0.001710         & 236.022156 $\pm$  0.002056         & 236.025192  $\pm$ 0.004281 \\
longitude of the ascending node [deg]      &  88.181668 $\pm$  0.000236         &  88.182090 $\pm$  0.000461         &  88.182237 $\pm$ 0.000973 \\
epoch of osculation [TT]                   &  2017\,07\,05                      & 2017\,07\,05                       & 2017\,07\,05 \\
 RMS                                       &  0.38 arcsec                       & 0.47 arcsec                        &  0.52 (normalized)   \\
 &  &  & \\
\hline 
\end{tabular}
}}
\end{table*}

\begin{table*}
\caption{Original and future barycentric orbits of K2 at 250\,au from the Sun, i.e. before entering and after leaving the planetary zone, respectively. Inverse semimajor axis values are in units of  $10^{-6}$au$^{-1}$, and angular orbital elements are given in degrees.}
\label{tab:orbit_origin_future}
{\scriptsize{
\setlength{\tabcolsep}{8pt} 
\begin{tabular}{lcccc}
\hline
\hline 
 &  &  & & \\
                                           & \multicolumn{2}{c}{\bf S o l u t i o n ~~~~~ A1}                        & \multicolumn{2}{c}{\bf S o l u t i o n ~~~~~ A0} \\
                                           
                                           & original orbit                     & future orbit                        & original orbit                     & future orbit                       \\
 &  &  & & \\
\hline 
 &  &  & & \\
perihelion distance [au]                   & 1.79551272 $\pm$ 0.00005475        & 1.79335881 $\pm$ 0.00005584        & 1.79546949 $\pm$ 0.00006519        & 1.79331479 $\pm$ 0.00006646        \\
eccentricity                               & 0.99991296 $\pm$ 0.00001423        & 0.99795491 $\pm$ 0.00001431        & 0.99990416 $\pm$ 0.00001821        & 0.99794611 $\pm$ 0.00001833        \\
inverse of the semimajor axis              & 48.48 $\pm$ 7.93          & 1140.4 $\pm$ 8.0                   & 53.38 $\pm$ 10.15                  & 1145.3 $\pm$ 10.3                   \\ 
time of perihelion passage [TT]            & 2022\,12\,18.855081 $\pm$ 0.029861 & 2022\,12\,19.872785 $\pm$ 0.030085 & 2022\,12\,18.862555 $\pm$ 0.041019 & 2022\,12\,19.880427 $\pm$ 0.041289 \\
inclination                                &  87.573197 $\pm$  0.000023         &  87.789103 $\pm$  0.000017         &  87.573129  $\pm$ 0.000045         &  87.789042 $\pm$  0.000042         \\
argument of perihelion                     & 236.214199 $\pm$  0.001702         & 236.051977 $\pm$  0.001691         & 236.215269  $\pm$ 0.002075         & 236.053032 $\pm$  0.002063         \\
longitude of the ascending node            &  88.086028 $\pm$  0.000236         &  88.254521 $\pm$  0.000240         &  88.086450  $\pm$ 0.000460         &  88.254950 $\pm$  0.000464         \\
epoch of osculation [TT]                   &  1722\,10\,20                      & 2336\,07\,13                       &  1722\,10\,20                      & 2336\,08\,22              \\
 &  &  & & \\
\hline 
\end{tabular}
}}
\end{table*}

\begin{table*}
\caption{Barycentric orbit elements of K2 at the previous and next perihelion for two considered  solutions A1 and A0. Inverse semimajor axis values are in units of  $10^{-6}$au$^{-1}$. 
Orbital elements are described either by a mean value (with a $1\sigma$ uncertainty) when a normal distribution is applicable, otherwise by three deciles at 10, 50 (median) and 90 per cent. }
\label{tab:orbit_previous_next}
{\scriptsize{
\setlength{\tabcolsep}{12.0pt} 
\begin{tabular}{lcccc}
\hline 
\hline
 &  &  & & \\
                                           & \multicolumn{2}{c}{\bf S o l u t i o n ~~~~~ A1}                        & \multicolumn{2}{c}{\bf S o l u t i o n ~~~~~ A0} \\
                                           & previous orbit                     & next orbit                          & previous orbit                     & next orbit       \\
 &  &  & & \\
\hline 
 &  &  & & \\
perihelion distance [au]                   & 2.625 : 3.815 : 8.335         & 1.793346 $\pm$ 0.000056            & 2.264 : 3.041 : 6.391        & 1.793300  $\pm$0.000067      \\
aphelion distance [$10^3$au]               & 34.22 : 41.45 : 52.44         & 1.752  $\pm$ 0.012                 & 30.11 : 37.40 : 48.98        & 1.744 $\pm$ 0.016            \\
inverse of the semimajor axis              & 48.28 $\pm$ 7.91              & 1140.2 $\pm$ 8.0                   & 53.48 $\pm$ 10.14            & 1145.4 $\pm$ 10.3            \\ 
estimated time of perihelion passage [$10^6$\,yr from now] & -4.22 : -2.97 : -2.26  & 0.03270 $\pm$ 0.00027     & -3.81 : -2.54 : -1.84        & 0.03252 $\pm$ 0.00035        \\
percentage  of dynamically old VCs (q<10\,au) & 97 per cent                   & 100 per cent                       & 97 per cent                  & 100 per cent                 \\
 &  &  & & \\
\hline 
\end{tabular}
}}
\end{table*}

\section{Observational material}
The observational material was retrieved on 2018 February 21 from the Minor Planet Center.
It consists of 450 positional measurements covering a period from  2013 May  12 to  2018 January 23. This time interval of data is  four months longer than in \citet{hui:2017}, which means, however, that it is longer only by 7 per cent. 
\newline The positional observations are distributed heterogeneously over the studied period.
K2 was discovered on 21 May 2017 by the Panoramic Survey Telescope and Rapid Response System (Pan-STARRS) in Hawaii when it was 16.1\,au from the Sun.Later on, almost twenty pre-discovery measurements dating back to 12 May 2013 were found. However, one large gap of about 2.5\,yr and two shorter of about six months and eight months still exist inside this pre-discovery interval of measurements. Since its discovery the comet has been observed continuously. The distribution of positional observations as well as heliocentric and geocentric distance changes of K2 are shown in Fig.~\ref{fig:data_dist}. It is worth to mention here, that such a range of astrometric cometary positions is almost optimal for the past dynamical history investigations due to almost five years of its length and the lack of non-gravitational effects at such a large heliocentric distance.

\section{\label{orbit:obs-ori-fut}Observed, original and future orbit of K2}

We determined the osculating orbit (hereafter nominal orbit) from all positional measurements available in the MPC database\footnote{\tt http://www.minorplanetcenter.net/db{\_}search} on 21~February 2018. The details about selection and weighting of positional data are described in \citet{kroli-dyb:2010}. 

We present here two solutions. The first one (marked as A1) is based on weighted data (second column of Table~\ref{tab:orbit_observed}) and we treated it as the preferred solution. For comparison, we also present the solution based on unweighted data (solution A0, third column of Table~\ref{tab:orbit_observed}) to show the magnitude of orbital element dispersions at the important  stages of a dynamical evolution of K2: at 250\,au from the Sun (original and future orbits), and at the moments of a previous and next perihelion (previous and next orbit). Both our solutions are consistent with each other within about 1$\sigma$ combined error and also with the JPL solution based on the same time interval (last column of Table~\ref{tab:orbit_observed}; it is important to stress that we use the different selection and weighting scheme for data than the JPL used). 

\noindent \citet{hui:2017} noticed that K2 activity has been slowly increasing since 2013. However, 
no one has ever been able to derive a reliable non-gravitational orbit for a comet that is so far away from the Sun, even assuming other forms of a g(r)-like function (see \citet{marsden-sek-ye:1973}) compatible with a CO-sublimation. We have checked that it is not possible in this case either (because the non-gravitational effects are too weak). 

Next, in order to control the uncertainties in the motion of K2 at each stage of our dynamical calculations, we cloned its orbit and constructed a swarm of 5\,000 VCs resembling the observations (using the method described by \citet{sitarski:1998}). 
This allowed us to present the uncertainties of all orbital elements for dynamically evolved orbits, i.e. original and future  orbits at 250\,au from the Sun when planetary perturbations are negligible. These orbits  are  presented in Table~\ref{tab:orbit_origin_future}. Values of $1/a_{\rm ori}$ for both our solutions are greater than that obtained by \citet{hui:2017}. 

At these steps of dynamical evolution of K2 the equations of motion have been integrated using the recurrent power series method \citep{sitarski:1989}, taking into account perturbations by all planets and including relativistic effects.

\section{\label{orbit:previous}K2 in its previous perihelion}

\begin{figure*}
	\includegraphics[angle=270,width=8.8cm]{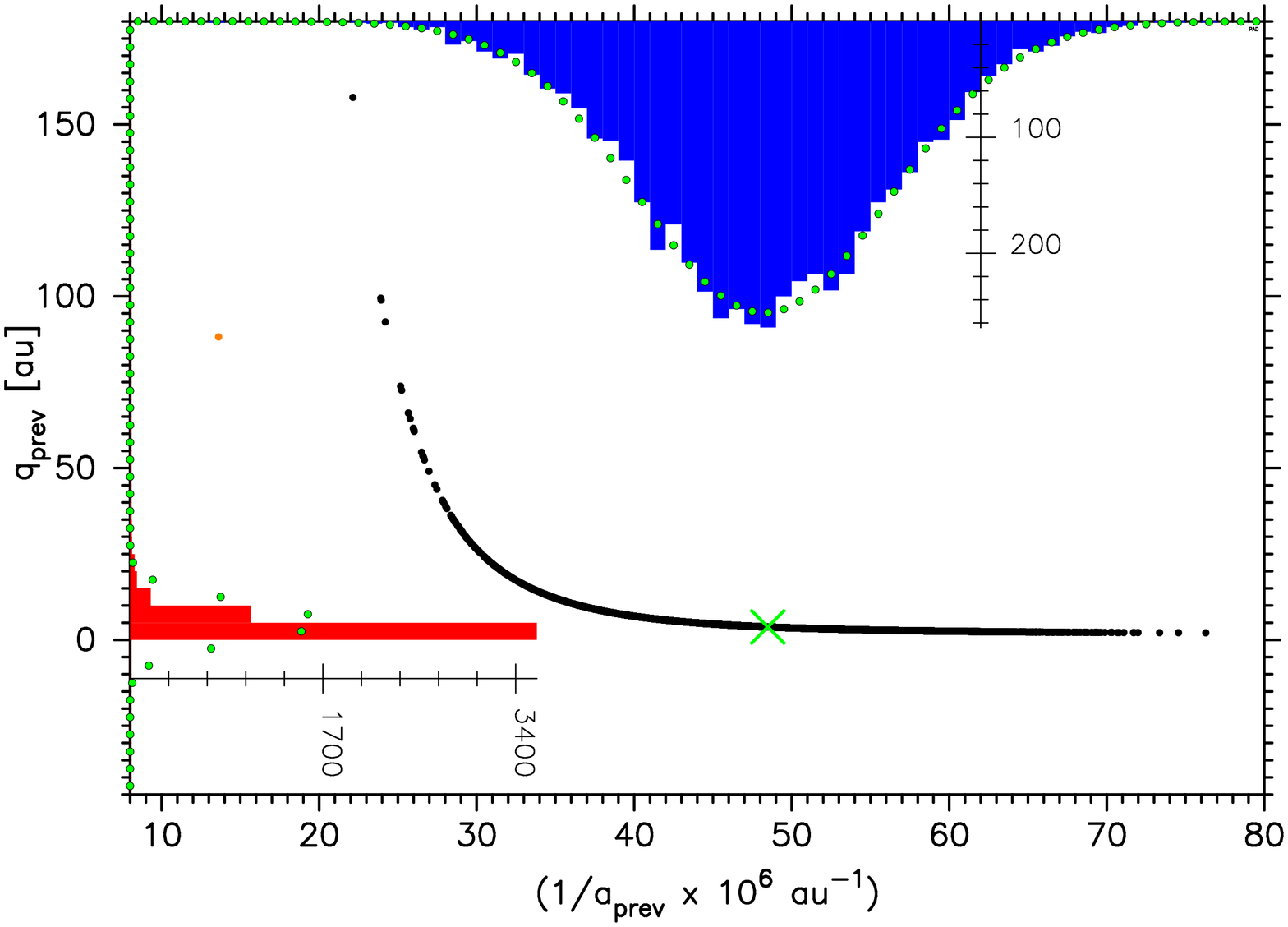} 
	\includegraphics[angle=270,width=8.8cm]{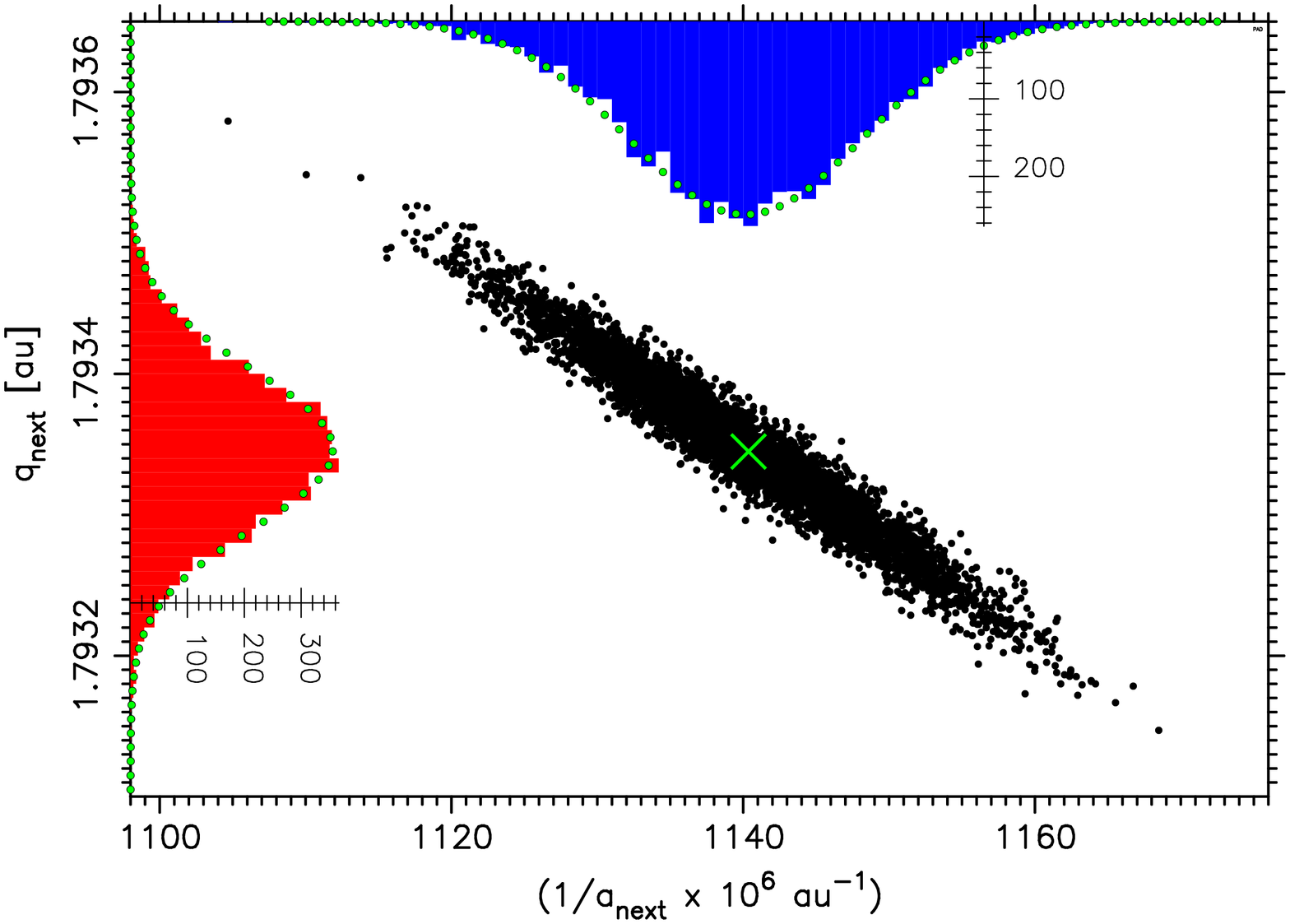} 
\protect\caption{\label{fig:A1_a_vs_q} Projection of 5\,001~VCs of K2
onto the 1/a--q plane (for the basic solution A1) augmented with two marginal distributions of these parameters. 
Left plot shows the orbital elements for K2 in the previous perihelion, right plot -- the orbital elements in the next perihelion.
In each picture, the black point depicting the evolved nominal orbit is situated in the centre of a green circle, and full green dots show a Gaussian fitting to $1/a$ and $q$ marginal distributions.
One can see that the distribution of $q_{\rm prev}$ (red histogram in the left picture) shows clear departures from a Gaussian distribution.} 
\label{fig:stat}

\end{figure*}

\begin{figure*}
	\begin{center}
		\includegraphics[angle=270,width=8.7cm]{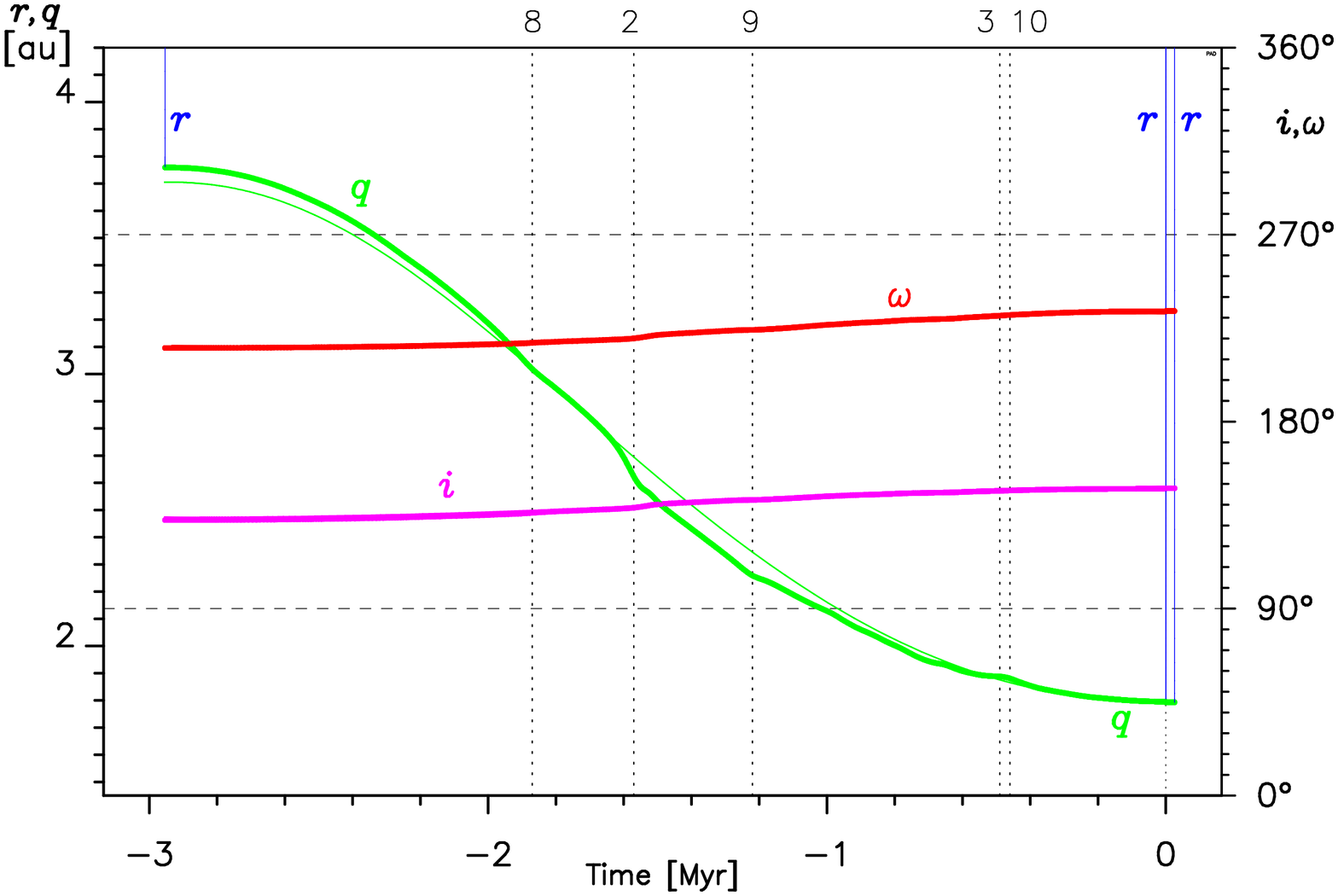} 
		\hspace{0.4 cm}
		\includegraphics[angle=270,width=8.7cm]{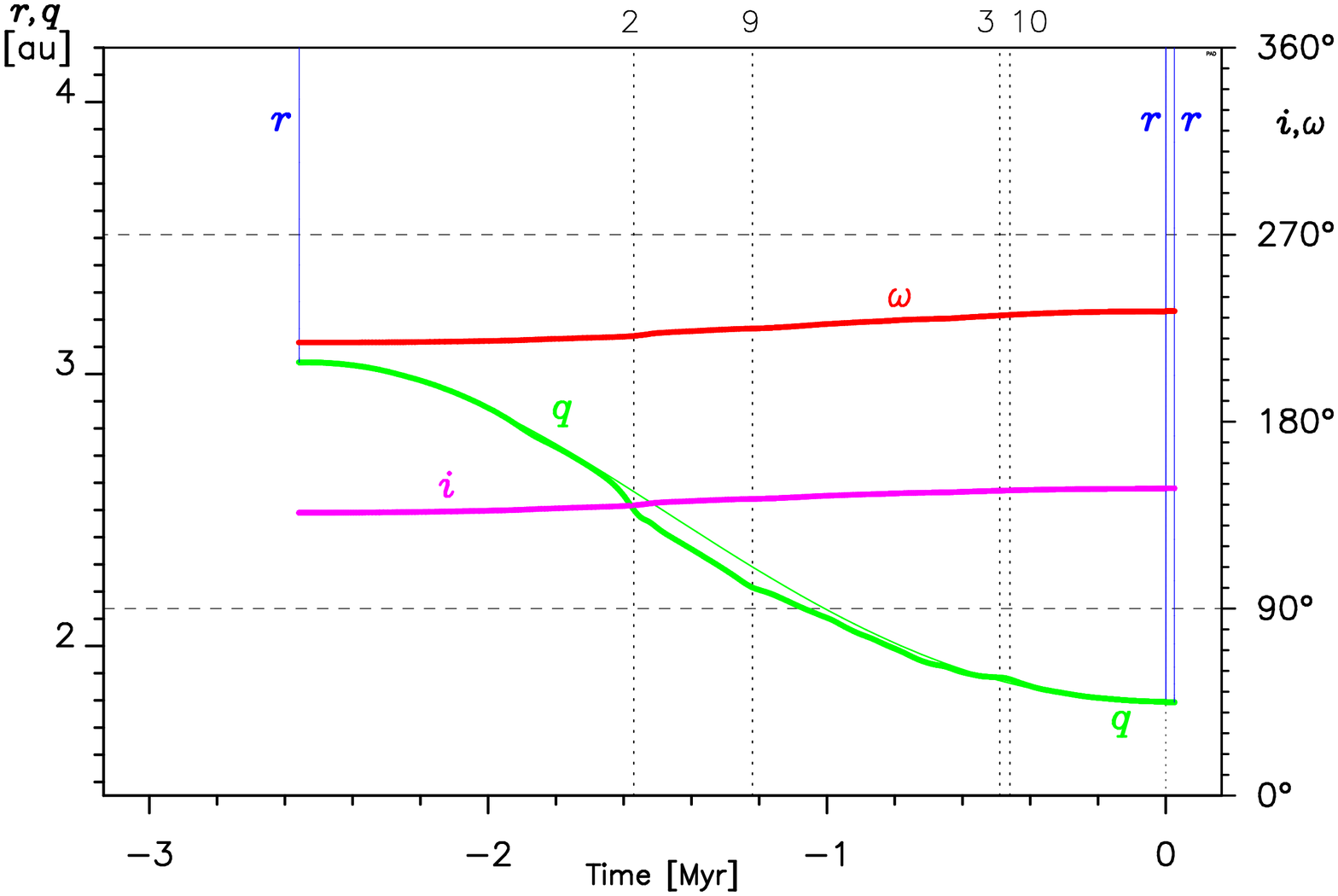}
	\end{center} 
	\caption{\label{fig:evol-gal-stars} Past and  future dynamics of K2 for solution A1 (left panel) and solution A0 (right panel). In both panels the horizontal time axis extends from the previous perihelion through the observed one up to the  next perihelion passage. See text for a detailed explanation.} 

\end{figure*}

At a distance beyond 250\,au from the Sun we can ignore planetary perturbations, and study the earlier dynamical evolution of K2 until it reaches its previous perihelion taking into account the perturbations from the Galactic disk, Galactic centre and all known potentially perturbing stars. The list of ten potentially strongest stellar perturbers of K2~past motion is presented in Table \ref{tab:stars}.

We use here the same model of Galactic gravitational potential as in \citet{kroli_dyb:2017}. Updated was only the list of known passing stars. This was a consequence of recently published results of Gaia \citep{GaiaDR1:2016} and RAVE \citep{RAVE-5:2017}. Ranges of selected orbital elements of K2 at the previous perihelion are shown in Table~\ref{tab:orbit_previous_next}, and the swarm of VCs projected onto the 1/a--q plane for the basic solution A1 is shown in the left panel of Fig.~\ref{fig:A1_a_vs_q}. One can see that a small percentage of VCs formed a dispersed wing stretching towards the upper left corner of this plot (large $q_{\rm prev}$, small $1/a_{\rm prev}$). This wing is so long that to keep a reasonable vertical scale we omit two VCs with extremely large values of $q_{\rm prev}$ (473\,au and 3037\,au, both at $1/a_{\rm prev}=22 \times 10^{-6}$au$^{-1} $). As a result of this dispersion the $q_{\rm prev}$-distribution is far from Gaussian. This basic solution (A1) gives the nominal previous perihelion of $q_{\rm prev} = 3.815$\,au and 97~per~cent of VCs with $q_{\rm prev}<10$\,au which means that we can conclude that K2 is a dynamically old comet which passed its previous perihelion about 3\,Myr ago deep inside the planetary zone.

\section{\label{orbit:next}K2 in the next perihelion}

We study the future motion of K2 to the next perihelion in a similar way as above. Orbital elements of K2 at the next perihelion are shown in Table~\ref{tab:orbit_previous_next}, whereas a full swarm of VCs projected onto the 1/a-q plane for preferred solution A1 is shown in the right panel of Fig.~\ref{fig:A1_a_vs_q}.
It turns out that the swarm of VCs at the next perihelion is very compact and gives precise values of orbital elements. According to our calculations K2 will pass next perihelion at the distance of about 1.8\,au, similarly as in the current passage, and its orbital period after leaving the planetary zone will be of only about 32\,500\,yr (Table~\ref{tab:orbit_previous_next}). 
We would like to stress, however, that the future dynamical evolution of K2 can be calculated in a certain way only after the comet passing through the perihelion, because non-gravitational effects can significantly change current predictions.

\begin{table}
	\caption{List of ten potentially strongest stellar perturbers of C/2017~K2 in order of their increasing comet--star minimal distances. For each star we present their estimated mass $M_{*}$ in solar masses, minimum distance $d_{min}$ between a comet and a star in parsecs, an epoch $T_{min}$ of this approach in Myrs and a relative velocity $V_{rel}$. Parameter $\gamma=M_{*}/(d_{min} \cdot V_{rel})$ is the theoretical approximate measure of the stellar perturbation strength but the actual perturbation depends also strongly on the geometry of the encounter.}
	
	\label{tab:stars}
	\begin{tabular}{ccccccc}
		\hline 
		\hline
		& Star name & $M_{*}$ & $\gamma$ & $d_{min}$ &  $T_{min}$ & $V_{rel}$ \\
		\hline
		1 &   GJ 217.1  &  2.4  & 0.239 &   1.31 & -0.85   &    7.65  \\
		2 &   HIP 30344            &  0.79 & 0.056 &   1.31 & -1.57   &   10.71  \\
		3 &   GJ 208   &  0.51 & 0.040 &   1.41 & -0.49   &    9.02  \\ 
		4 &   HD 49015  &  1.4  & 0.029 &   1.55 & -1.52   &   31.05  \\ 
		5 &   GJ 1049   &  0.51 & 0.016 &   1.56 & -0.62   &   20.34  \\ 
		6 &   HD 43947   &  1.2  & 0.026 &   1.75 & -0.66   &   26.41  \\ 
		7 &   HIP 27887            &  0.74 & 0.014 &   1.98 & -0.40   &   26.16  \\ 
		8 &   HD 37594 &  1.7  & 0.117 &   2.02 & -1.87   &    7.19  \\ 
		9 &   HD 54958  &  1.5  & 0.032 &   2.07 & -1.22   &   22.84  \\ 
		10 & GJ 120.1ABC    &  1.48 & 0.022 &   2.12 & -0.47   &   31.45  \\ 
		\hline 
	\end{tabular}
\end{table}

\section{\label{overall_evolution}Overall evolution during three perihelion passages}

The continuous dynamical evolution of K2 starting at the previous perihelion, through the observed one, up to the next perihelion passage is depicted in Fig.\ref{fig:evol-gal-stars}. In both panels the 	left vertical axis is expressed in au and corresponds to the   perihelion distance plot ($q$, green lines) as well as the heliocentric distance plots ($r$, thin, vertical blue lines). The right vertical axis is expressed in degrees and describes the evolution of the osculating inclination ($i$, magenta lines) and the argument of perihelion ($\omega$, red lines). Due to the invisible differences in $i$ and $\omega$ thin and thick lines overlap. Both these angular elements are 	expressed in a Galactic reference frame. Horizontal dashed lines call attention to the beginning of the second and fourth quarter of $\omega$, whose values (90\degr and 270\degr) are important from the point of view of the Galactic disk perturbations. Vertical dotted lines mark the epochs of the closest comet--star encounters with star numbers from Table \ref{tab:stars}  at the top of the panels. Not all stars from Table \ref{tab:stars} are marked because the actual perturbation depends strongly also on the geometry of the approach.

In both figures we compare the results obtained from a full force model (Galactic and stellar perturbations, thick lines) with a simplified one, where stellar perturbations are omitted (thin lines). 

It can be seen in Fig.~\ref{fig:evol-gal-stars} that K2 is now observed during its slowly decreasing phase of a perihelion distance evolution under the influence of a Galactic tide. In both A1 and A0 our solutions its most probable aphelion distance is rather small (41\,500\,au and 37\,400\,au, respectively) which makes this comet practically immune to any stellar perturbations. One can observe only small traces of interaction with HIP\,30344, HD\,54958, GJ\,208 and GJ\,120.1\,ABC stellar system during the last orbital period. Moreover, our extended calculations showed that, provided no significant planetary perturbations occurred during the previous perihelion passage, also its last but one approach to the Sun 6\,Myr ago took place at a small heliocentric distance of a few au.

According to current data K2 will be significantly perturbed by planets and as a result its semimajor axis will be shortened down to about 875\,au which completely switch off Galactic and stellar perturbations during its future orbital period lasting about 32\,500\,yr. Please keep in mind that non-gravitational effects due to a cometary activity near the perihelion as well as violent changes in its behaviour can modify this prediction. 

\begin{acknowledgements}
We wish to thank an anonymous referee for valuable comments and suggestions. This research was partially supported by the project 2015/17/B/ST9/01790 founded by the National Science Centre in Poland. 
\end{acknowledgements}

\bibliographystyle{aa}
\bibliography{moja24}

\vfill\eject
\onecolumn{

\begin{center}
\Large{\bf November 2018 Addendum}
\end{center}

\section{Summary}

The updated orbit for C/2017~K2 PANSTARRS is discussed here. This comet was in mid-October about 13\,au from the Sun and is expected to appear at perihelion in December 2022 ($q=1.81$\,au), so for the next few years all positional observations are of great importance for studying its past motion. It seems that now (November 2018) dynamical status of C/2017~K2 is still not fixed. According to dynamical evolution based on positional measurements taken between 2013 May~12 -- 2018 October~13 this comet passed the previous perihelion in a distance between 7\,au and 25\,au from the Sun.

\section{Updated orbital solutions for C/2017~K2}

We updated the orbital solution for C/2017~K2 (hereafter K2) on the basis of all positional measurements available in the MPC database\footnote{\tt http://www.minorplanetcenter.net/db{\_}search} on November the 5$^{th}$, 2018.
It means that new solutions were derived using a total of 996 observations spanning the time interval: 2013 May~12 -- 2018 October~13. Applied was a similar procedure as 
above (sections~\ref{orbit:obs-ori-fut}--\ref{orbit:previous}) and two osculating orbits: A0-new  (unweighted data) and A1-new  (weighted data) and respective original barycentric orbits are presented in Table~\ref{tab:orbit_observed_origin}. 
The observational material is significantly richer, so uncertainties of orbital elements are much smaller than before; compare Tables~\ref{tab:orbit_observed}--\ref{tab:orbit_origin_future} with the present Table~\ref{tab:orbit_observed_origin}.

\begin{table*}[h]
	\caption{Osculating heliocentric orbit and original barycentric orbit of K2 based on 996 positional measurements spanning the time interval from 2013 May~12 to 2018 October~13 available at MPC on 5 November 2018. Ecliptic of J2000 is used. Solution A1 is based on weighted measurements whereas A0 is based on unweighted data. Inverse semimajor axis values are in units of  $10^{-6}$au$^{-1}$, and angular orbital elements are given in degrees.}
	\label{tab:orbit_observed_origin}
	{\scriptsize{
			\setlength{\tabcolsep}{8pt} 
			\begin{tabular}{lcccc}
				\hline 
				&  &  & & \\
				&  \multicolumn{2}{c}{\bf O s c u l a t i n g ~~~~o r b i t}  &  \multicolumn{2}{c}{\bf O r i g i n a l ~~o r b i t}  \\
				&  Solution  A0-new                  & Solution A1-new                  &   Solution  A0-new & Solution A1-new  \\
				&  &  & & \\
				\hline 
				&  &  & & \\
				perihelion distance [au]                   & 1.80556190 $\pm$ 0.00001457        & 1.80554795 $\pm$ 0.00001000        & 1.79562804 $\pm$ 0.00001438 & 1.79561418 $\pm$ 0.00001007 \\
				eccentricity                               & 1.00036461 $\pm$ 0.00000612        & 1.00036331 $\pm$ 0.00000418        & 0.99993756 $\pm$ 0.00000614 & 0.99993627 $\pm$ 0.00000417 \\
				time of perihelion passage [TT]            & 2022\,12\,21.412478 $\pm$ 0.030195 & 2022\,12\,20.658440 $\pm$ 0.009958 & 2022\,12\,18.799681 $\pm$ 0.014635 & 2022\,12\,18.806851 $\pm$ 0.009899  \\
				inclination                          &  87.544002 $\pm$  0.000021         &  87.544065 $\pm$  0.000014         &  87.573087  $\pm$ 0.000021  &  87.573150  $\pm$ 0.000014 \\
				argument of perihelion               & 236.079112 $\pm$  0.000544         & 236.079444 $\pm$  0.000378         & 236.210878  $\pm$ 0.000547 &  236.211209  $\pm$ 0.000380 \\
				longitude of the ascending node      &  88.257353 $\pm$  0.000163         &  88.257072 $\pm$  0.000109         &   88.086869 $\pm$ 0.000166  &  88.086586 $\pm$ 0.000110 \\
				epoch of osculation [TT]                   &  2018\,11\,18                      & 2018\,11\,18                       & 1722\,12\,20 &  1722\,12\,20  \\
				RMS (No of residuals)                      &  0.53 arcsec (1972)                & 0.39 arcsec (1969)                 &             
				&    \\
				inverse of the semimajor axis              & $-$201.94 $\pm$ 3.39               & $-$201.22 $\pm$ 2.31                   & 34.78 $\pm$ 3.42            & 35.49 $\pm$ 2.32            \\ 
				&  &  & & \\
				\hline 
			\end{tabular}
	}}
\end{table*}

\begin{table*}[h]
	\centering
	\caption{Barycentric orbit elements of K2 at the previous perihelion for two considered  solutions A0n and A1n. Inverse semimajor axis values are in units of  $10^{-6}$au$^{-1}$. 
		Orbital elements are described either by a mean value (with a $1\sigma$ uncertainty) when a normal distribution is applicable, otherwise by three deciles at 10, 50 (median) and 90 per cent. }
	\label{tab:orbit_previous_new}
	{\small{
			\setlength{\tabcolsep}{12.0pt} 
			\begin{tabular}{lcc}
				\hline 
				&  &   \\
				& \multicolumn{1}{c}{\bf S o l u t i o n ~~~~~ A0-new} & \multicolumn{1}{c}{\bf S o l u t i o n ~~~~~ A1-new} \\
				& previous orbit               & previous orbit           \\
				&  & \\
				\hline 
				&  &  \\
				perihelion distance [au]                   & 7.54 : 12.61 : 24.96          & 8.04 : 11.41 : 17.23            \\
				aphelion distance [$10^3$au]               & 51.16 : 57.61 : 65.95         & 51.99 : 56.37 : 61.42           \\
				inverse of the semimajor axis              & 34.75 $\pm$ 3.42              & 35.50 $\pm$ 2.33                \\ 
				estimated time of perihelion passage [$10^6$\,yr from now] & -5.93 : -4.85 : -4.07  & -5.34 : -4.70 : -4.16  \\
				percentage  of dynamically old VCs ($q$<10\,au) & 30 per cent                   & 33 per cent               \\
				percentage  of  VCs with 10\,au$<q<$15\,au      & 34 per cent                   & 48 per cent               \\
				percentage  of dynamically new VCs ($q$>20\,au) & 18 per cent                   & ~5 per cent               \\
				&  &  \\
				\hline 
			\end{tabular}
	}}
\end{table*}

From a point of view of the past motion of K2 the most important change is the significant increase of its original semimajor axis, from previous 20\,600~au to current 28\,000~au. This makes K2 more sensitive to Galactic and stellar perturbations which causes its swarm of VCs to be more dispersed at a previous perihelion.

\begin{figure}
	\centering
	\includegraphics[angle=270,width=12.5cm]{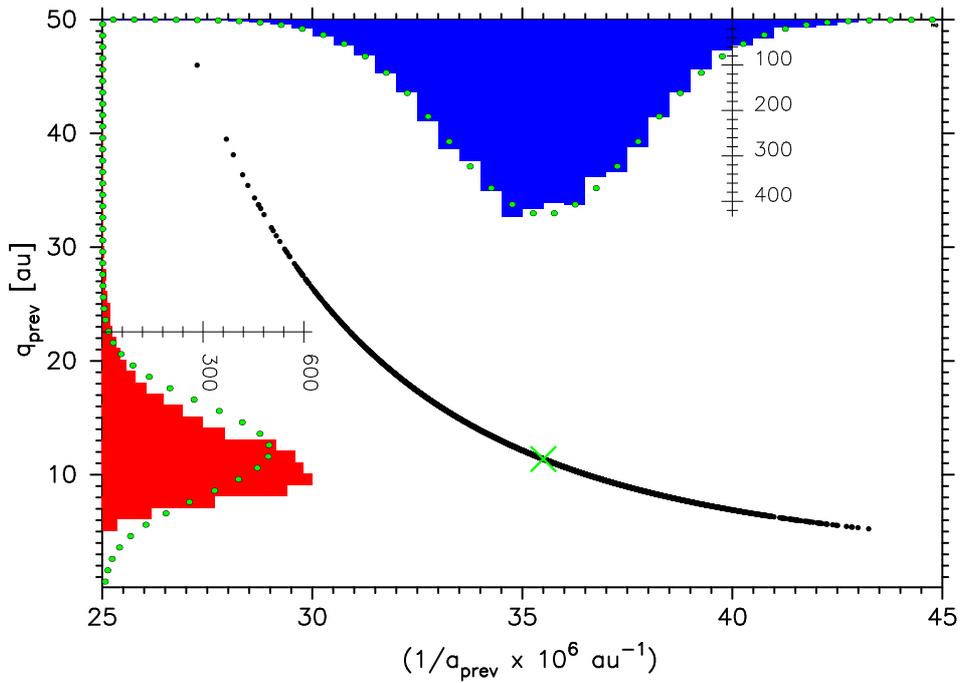} 
	\caption{\label{fig:A1_a_vs_q} Projection of 5\,001~VCs of K2 onto the 1/a--q plane for the solution A1-new augmented with two marginal distributions of these parameters. This is an updated version of the left part of Fig.~\ref{fig:stat}.}
\end{figure}

\section{K2 at its previous perihelion}

Using exactly the same model of Galactic potential and a list of potentially perturbing stars as in Section~\ref{orbit:previous}, we obtained new results for orbital parameters in the previous perihelion, see Table~\ref{tab:orbit_previous_new}. One can see that now a swarm of VCs stopped at the previous perihelion is more scattered  because now  Galactic perturbations act stronger and during a longer orbital period of about 5 Myr.  As a result, the dynamical status of K2 is less certain than before. According to a weighted solution A1-new only 33~per cent of the swarm of VCs passed within 10\,au from the Sun during the previous perihelion passage, however still only 5~per cent of them passed further than 20\,au from the Sun. The two-dimensional distribution of this solution swarm stopped at the previous perihelion is presented in Fig.~\ref{fig:A1_a_vs_q}. The previous perihelion distribution is still non-Gaussian so in Table~\ref{tab:orbit_previous_new} we present it (and the aphelion distance) as the 10$^{th}$ decile, a median and the 90$^{th}$ decile as in Table~\ref{tab:orbit_previous_next}. 

\section{Conclusion}
From the results presented in this update note it should be concluded that the dynamical status of K2 is not as obvious as it seemed to us earlier. We plan to prepare  \footnote{ at {\tt ssdp.cbk.waw.pl/LPCs} or at our new cometary database which will start in 2019} the next update in the course of C/2017~K2 approaching its perihelion in December 2022 or a little earlier, when it will be about 5\,au from the Sun (to avoid problem with non-gravitational effects). One should expect further decreasing of the uncertainties of osculating orbital elements which might result also in a more precise previous orbit perihelion distance.

}

\end{document}